\documentclass[10pt,journal]{IEEEtran}
\usepackage{graphicx}
\usepackage{amsmath,amssymb,amsfonts}
\usepackage{booktabs,array,calc,tikz}
\usepackage{cite}
\usepackage{url}
\usepackage[colorlinks=true,allcolors=blue]{hyperref}
\providecommand{\tightlist}{\setlength{\itemsep}{0pt}\setlength{\parskip}{0pt}}
\providecommand{\real}[1]{#1}
\begin{document}
\title{Decoding Motion-Blurred QR Codes: Motion-Parameter Error Tolerance and Model-Based Recovery}
\author{Zhi~Zeng\textsuperscript{*},~\IEEEmembership{Member,~IEEE,} Qiaochen~Su, Si~Xiong, Yaoqi~Wang, Yufei~Cai, Heng~Liu, and~Yi~Hu%
\thanks{This work was supported in part by the National Natural Science Foundation of China under Grant 61805185 and in part by the Key Core Technology Research and Development Program of Taiyuan under Grant 2025TYJB02.}%
\thanks{Zhi Zeng is with the State Key Laboratory of Electromechanical Integrated Manufacturing of High-Performance Electronic Equipment, Xidian University, Xi'an 710071, China, and also with the School of Mechano-electronics Engineering, Xidian University, Xi'an 710071, China. \textsuperscript{*}\emph{Corresponding author: Zhi Zeng} (e-mail: zhizeng@mail.xidian.edu.cn). Qiaochen Su, Si Xiong, Yaoqi Wang, Yufei Cai, Heng Liu, and Yi Hu are with the School of Mechano-electronics Engineering, Xidian University, Xi'an 710071, China.}}
\maketitle
\begin{abstract}
In automated manufacturing, logistics, and inspection, Quick Response (QR) codes are often read on the fly under continuous ambient light without active strobes or motion pauses. The resulting relative three-dimensional motion produces spatially varying blur that defeats standard decoders, yet how accurately the motion must be estimated to recover the payload remains an open question. To address this question from a task-driven evaluation perspective centered on machine recognition effectiveness, we formulate motion-blurred QR recovery within a unified mathematical framework of identifiability. Within this framework, we systematically measure the decoding tolerance to perturbations in motion length, direction, rotation or scaling centre, and speed profile. These controlled experiments reveal three structural laws: 1) parameter tolerances shrink rapidly as blur amplitude grows and are tightest under uniform speed, whereas accelerating motion is substantially more forgiving; 2) length tolerance can differ markedly between underestimation and overestimation; 3) errors that are individually benign interact strongly and prevent decoding when combined. Guided by these findings, we propose a grey-box recovery pipeline in which two compact networks predict an initial corner trajectory, followed by tolerance-guided candidate search, alternating image-and-path fitting, and regularised linear inversion. On a benchmark of 955 real motion-blurred photographs spanning four camera models and four rigid-motion families, our method recovers the exact payload on 93.1\% of the images. It outperforms the strongest fine-tuned deblurring network at 83.6\% and reduces the decoding error rate by 58.0\%, with zero wrong strings accepted.
\end{abstract}
\section{Introduction}\label{introduction}

Quick Response (QR) codes serve as a foundational link between physical parts and digital identity across manufacturing, logistics, and mobile inspection. Outside fixed-speed scan tunnels, codes are routinely captured by general-purpose cameras on reconfigurable workcells, robotic handlers, or mobile readers. To keep the workflow compact and maintain throughput, neither the moving part nor the camera can pause or slow down for each read. At the same time, active strobe illumination is often undesirable in these settings, so the camera must operate under continuous passive ambient light. Consequently, during a single exposure, continuous relative motion combines lateral travel, axial distance change, rotation, and acceleration into spatially varying motion blur across the planar code.

When motion blur smears each module into its neighbours, a decoder can no longer locate a consistent grid. Once too many modules are corrupted, built-in Reed--Solomon error correction \cite{iso18004} also fails to recover the payload. In automated tracking, an unread frame halts the automatic flow and returns the item to manual handling, whereas a misread payload sends the item forward under a false identity. Restoring the blurred photograph before decoding is therefore a natural remedy, and its operational goal is to maximise exact payload recovery without accepting any wrong string.

\begin{figure*}
\centering
\centering\includegraphics[width=0.90\linewidth]{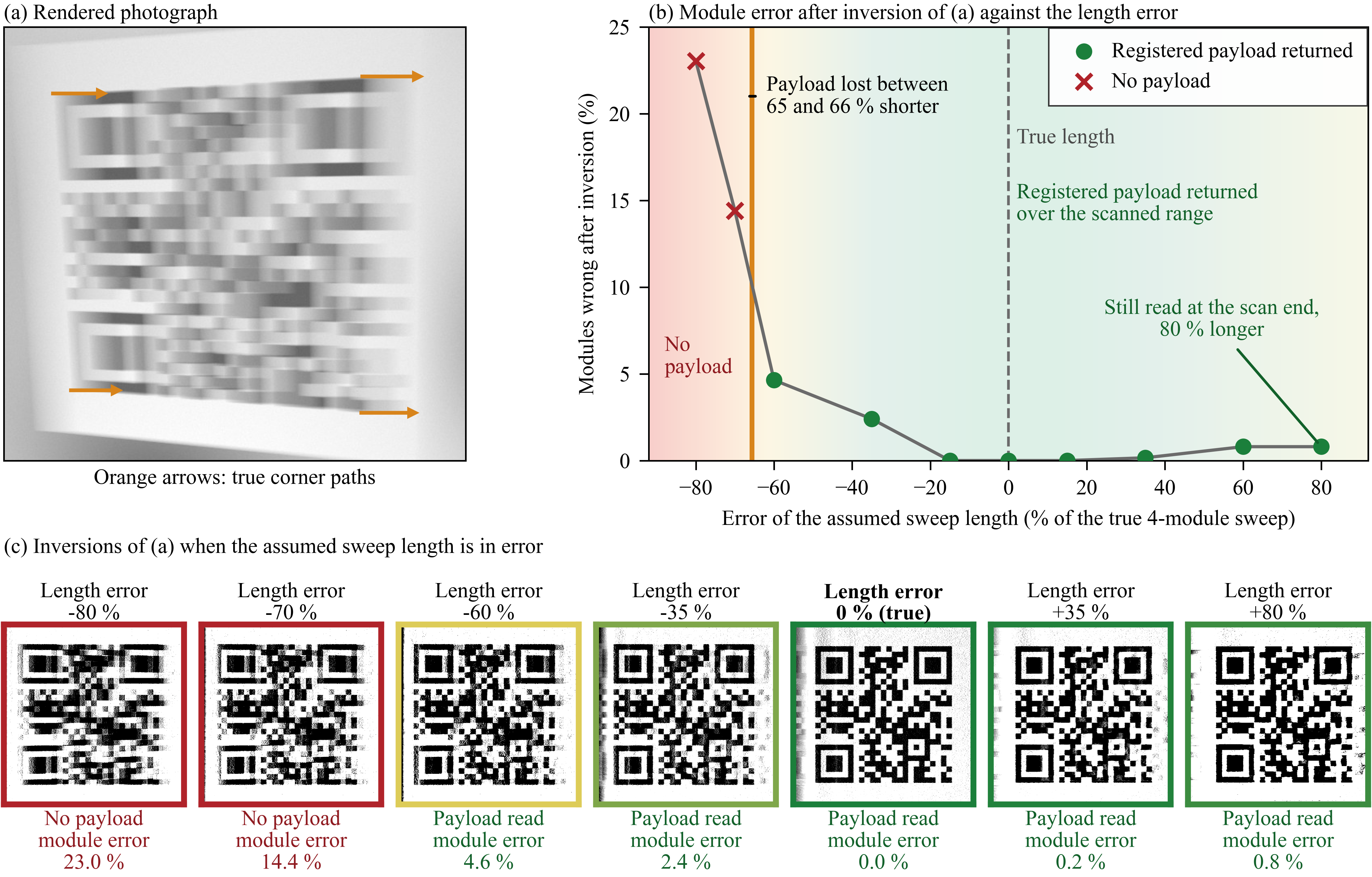}
\caption{An illustration of how motion-estimation error affects the module error rate and decoding outcome. As seen in (b) and (c), the accuracy of the motion estimate directly governs deblurring accuracy and decoding success, while the error tolerance of this motion parameter can be markedly asymmetric. \textbf{(a)} The rendered photograph of a four-module lateral sweep whose speed rises 5.7-fold, with orange arrows marking the true corner paths. \textbf{(b)} Fraction of modules recovered wrongly when the same photograph is inverted with the assumed sweep length in error. Green: the three-decoder read-out returns the registered payload; red: no payload. The payload is lost between 65 and 66\% too short and survives at 80\% too long, the end of the scan. \textbf{(c)} The inversions behind seven of the points in (b). Frame colour: red, no payload; yellow to green, module error falling from 5\% to 0\%.}
\end{figure*}

To restore a blurred code, existing methods draw on three complementary sources of information: the known structure of the code, the physics of image formation, and statistical priors learned from data. Because the finder patterns and binary module grid are known in advance, several methods estimate a single blur kernel directly from them \cite{esedoglu2004,choksi2010,vangennip2015,zheng2024,rioux2021}. For a distant or planar scene, camera motion during exposure traces a single low-dimensional trajectory \cite{gupta2010,whyte2012,tai2011}, which tightly constrains the resulting blur operator \cite{levin2011}. Learned methods either predict an explicit degradation or map the blurred image directly to a sharp output. The former approach predicts a pixel-wise kernel field \cite{sun2015,gong2017,carbajal2023}, low-dimensional kernel parameters \cite{fang2025} or a camera trajectory \cite{carbajal2025}, whereas the latter ranges from multi-scale convolutional networks \cite{nah2017} to transformer restorers \cite{restormer2022,tsai2022}. For QR codes, Shi et al.~follow the explicit parameter approach \cite{shi2025} and ADNet follows the direct restoration approach \cite{adnet2025}.

All three routes hinge on a common requirement: the underlying motion must be estimated accurately enough, explicitly or implicitly, for the restored code to decode \cite{vangennip2015,adnet2025}. Blur2Seq addresses this sensitivity by refining its predicted trajectory against the observed image and training its restorer to tolerate residual trajectory error \cite{carbajal2025}.

What remains unknown is the quantitative tolerance: how accurate must the motion estimate be? Overall decoding rate evaluates the entire pipeline at once and cannot separate geometric errors from kernel or noise errors. Related studies have examined parameter mismatch for one-dimensional barcodes \cite{iwen2013} and evaluated kernel perturbations on natural images using PSNR and SSIM \cite{ji2012,ren2018,nan2020,tang2023,motorcu2026,boracchi2012}. Yet neither answers, for a two-dimensional QR code under rigid-plane motion, how far each assumed motion parameter may depart from the truth before standard decoders fail to recover the payload. Fig. 1 illustrates this question on a single rendered photograph. In Fig. 1a, the card sweeps sideways by four modules while its speed increases 5.7-fold. In Fig. 1b and 1c, the blurred photograph is held fixed while only the assumed sweep length is varied prior to inversion and decoding. Decoding fails when the assumed sweep is 65--66\% too short, yet succeeds even when it is 80\% too long. Thus, decoding succeeds whenever the assumed motion lies within a tolerance basin around the truth, and both the size of that basin and how to reach it are the focus of this paper.

We can measure these motion-parameter tolerances through a task-driven image quality assessment centered on machine recognition effectiveness. For a given decoder and a known payload, a restored image either decodes or fails. Decodability can therefore be traced directly as a function of each parameter perturbation, rather than inferred indirectly from pixel-level perceptual metrics. A tight tolerance calls for precise estimation and refinement, whereas a wide tolerance does not. Following this principle, our study proceeds in four steps.

First, we formulate a mathematical framework that specifies which motions our model covers and when the resulting blur can be inverted. We express motion blur as a linear exposure operator determined by an invertible coordinate mapping and a one-dimensional convolution kernel \cite{robbins1972,sawchuk1974}: the camera trajectory induces the geometric mapping, and the exposure time spent along the path defines the kernel. The coordinate mapping itself is lossless; information is lost only where the convolution kernel attenuates spatial frequencies below the noise floor. Consequently, regularised linear inversion can recover the remaining frequencies without a learned image prior.

Second, we measure parameter tolerances through controlled perturbations. Holding each rendered photograph fixed, we perturb one motion parameter of the assumed operator away from its true value while keeping all other parameters at truth. A fixed linear inverter and standard decoders process the result, and bisection brackets the failure boundary. Across 1,104 synthetic photographs spanning lateral translation, axial approach, roll and tilt, we perturb blur length, direction, speed profile, and motion centre or axis, expressing each tolerance in the parameter's physical unit.

Third, guided by these measured tolerances, we design an explicit grey-box recovery pipeline. If direct decoding of the input photograph fails, two compact networks predict the code corners and an initial quadratic corner trajectory across the exposure. Because blur length and direction exhibit tight module-scale tolerances, a neighbourhood search refines both parameters, ranking candidates by reconstruction consistency on the fixed QR function patterns. Three rounds of alternating optimization then refine the binary module estimate and refit the corner trajectory to the photograph, followed by regularised linear inversion. To prevent false reads, a payload is accepted only when at least two of the three decoders return the identical string.

Finally, we validate the complete pipeline on a benchmark of 955 real motion-blurred photographs of printed QR cards captured across four camera models. Leveraging the equivalence between camera and target relative motion, controlled operator-driven sweeps span lateral translation, axial approach or retreat, in-plane rotation, and coupled 3D motion, with paired static references. Against 12 fine-tuned restoration networks and two classical methods, our pipeline lowers the decoding error rate from 16.4\% to 6.9\%, a 58.0\% relative reduction, with zero wrong strings accepted.

\textbf{Contributions.}

\begin{itemize}
\tightlist
\item
  \textbf{A mathematical framework linking motion-parameter error to payload recovery.} Specifically, motion blur is defined by an invertible image mapping and a convolution kernel. We state when such a mapping exists and what a regularised linear inversion can recover.
\item
  \textbf{Quantitative laws of motion-estimation tolerance.} For a QR code on a rigid plane, we quantify how far the assumed length, direction, centre or axis, and speed profile may depart from truth before decoding fails. This reveals how the tolerance depends on the speed profile, differs between underestimation and overestimation, and tightens under combined errors.
\item
  \textbf{A tolerance-guided grey-box recovery model.} It consists of two compact networks, a parameter search, and alternating fits to estimate the motion, together with a regularised linear inversion to restore the image. Every stage and decision can be inspected.
\item
  \textbf{A multi-motion, multi-device real-photograph benchmark.} Across four camera models and five operators, 955 real motion-blurred photographs with paired static references cover lateral translation, approach or retreat, rotation, and coupled 3D motion. The accompanying evaluation protocol strictly separates accepted wrong strings from rejections.
\end{itemize}

\section{Related Work}\label{related-work}

Prior methods for barcode and QR-code deblurring largely rely on known symbol priors. Van Gennip et al.~recover a single blur kernel by matching the known top-left finder pattern \cite{vangennip2015}, while Zheng et al.~estimate region-wise kernels using priors on both finder and alignment patterns \cite{zheng2024}. Other QR deblurring methods assume a known motion direction or guide multi-scale estimation with a sharpness metric \cite{shi2020,chen2021}. For general binary symbol fields, Rioux et al.~alternate between image and kernel estimation and test how large a kernel width their method can handle before decoding fails \cite{rioux2021,rioux2021ip}. For one-dimensional barcodes, Iwen et al.~bound decoding error under a misspecified Gaussian kernel width \cite{iwen2013}, and Sörös et al.~alternate edge-based kernel estimation with ZBar decoding until a checksum passes \cite{soros2015}. However, these methods assume spatially invariant kernels and do not systematically characterise how multi-parameter motion errors affect two-dimensional code decoding.

To handle spatially varying blur, camera motion during exposure has been modelled as a motion density or trajectory \cite{gupta2010,whyte2012,tai2011}, or assisted by inertial measurements \cite{zhang2016tci}. Whereas the rotational model of Whyte et al.~assumes a distant scene, Rao, Purohit and Rajagopalan model a planar target under general camera motion through a family of homographies \cite{rao2014,purohit2022}. Classically, Robbins and Huang \cite{robbins1972} and Sawchuk \cite{sawchuk1973,sawchuk1974} showed that certain spatially varying blurs reduce to standard convolution after an invertible coordinate transformation, a property recently applied to rotary blur in polar coordinates \cite{qin2025}. Building on this principle, we establish the exact reduction conditions for a rigid planar code and quantify the resulting motion-parameter tolerances.

Another line of work uses deep networks to predict explicit degradation parameters. For natural images, J-MKPD predicts a pixel-wise kernel field, and Blur2Seq regresses a three-dimensional camera-rotation trajectory \cite{carbajal2023,carbajal2025}, while algorithm-unrolling networks embed the forward blur operator across iterative restoration stages \cite{li2020tci,sanghvi2022tci}. Fang et al.~estimate local kernel length, orientation and curvature \cite{fang2025}, while for QR codes Shi et al.~place kernel estimation and Wiener filtering ahead of a transformer restorer \cite{shi2025}. Our networks likewise predict explicit geometric parameters---specifically, a quadratic corner trajectory covering translation, axial distance change, roll and tilt---which then serves as an initialization for candidate search and alternating fitting.

Several studies have examined image restoration under an inaccurate blur operator. Ji and Wang explicitly modelled kernel errors during deconvolution \cite{ji2012}, and Ren et al.~identified the frequency components that remain reliable under an inaccurate kernel \cite{ren2018}. Nan and Ji learned to handle kernel uncertainty \cite{nan2020}, while Mai and Liu fused kernels from multiple estimators \cite{mai2015}. Closer to our experimental design, Tang et al.~perturbed kernel width, length or orientation to trace PSNR and SSIM degradation \cite{tang2023}, and Motorcu and Çetin rotated the true kernel by up to 60° \cite{motorcu2026}. While our analysis also pairs a fixed blurred observation with a perturbed operator, we evaluate machine recognition effectiveness rather than pixel-level fidelity. Specifically, we perturb physical motion parameters in their native units and measure whether standard decoders recover the exact payload.

In contrast to model-based inversion, end-to-end restoration networks map blurred images directly to sharp outputs, evolving from multi-scale convolutional architectures \cite{nah2017,cho2021,liu2022tmm} to transformers, idempotent and degradation-aware baselines \cite{restormer2022,tsai2022,chen2022,mao2023tcsvt,song2024pr,jiang2024tim} as well as mixture-of-experts decoders \cite{feijoo2025}. QR-specific restoration networks include adversarial generators \cite{wang2019,dong2024}, a lightweight DeblurGAN-v2 variant \cite{gu2024}, edge-guided multi-scale networks \cite{egmsnet2025} and causal state-space models \cite{lcmqrir2026}. ADNet further incorporates a decoder into the inference loop, routing mild blur to a lightweight network and severe blur to EG-Restormer \cite{adnet2025}.

Finally, existing QR deblurring studies typically evaluate recovery with ZBar alone \cite{vangennip2015,adnet2025}, while EGMSNet reports ZBar and WeChat separately \cite{egmsnet2025}. By contrast, we require agreement across three independent decoders and verify the accepted string against an independently registered reference payload. In addition, Skurowski et al.~showed that reliable QR detection requires roughly 3 to 3.5 pixels per module \cite{skurowski2022}; we treat adequate spatial sampling as a precondition and render synthetic codes at eight or more pixels per module.

\section{Framework}\label{framework}

We describe motion blur within a unified exposure framework that covers every blur reducible to a convolution under an invertible change of coordinates. The framework separates the geometric motion mapping from the convolution kernel, and characterises both what information is lost and what regularised linear inversion recovers. We then apply it to short exposures of a rigid planar code, where the motion is determined by the trajectories of the four code corners. Supporting derivations for the convolution reduction, the frequency-domain inversion and the planar homography generator are given in Section S1 of the supplementary material.

\subsection{The exposure operator}\label{the-exposure-operator}

Let us begin with the exposure operator. Let \(\Omega\subset\mathbb{R}^2\) be the reference image region, in pixel units. The exposure lasts \(T>0\) seconds, and \(t_{\mathrm{ref}}\in[0,T]\) denotes the reference instant. The sharp image \(I_{\mathrm{s}}:\Omega\to[0,1]\) gives the normalised linear irradiance at each position at \(t_{\mathrm{ref}}\). Let \(\mathbf{g}(\mathbf{p}_0,t)\in\mathbb{R}^2\) be the image position at time \(t\) of the scene point located at \(\mathbf{p}_0\in\Omega\) at \(t_{\mathrm{ref}}\). For each \(t\), the mapping \(\mathbf{g}(\cdot,t)\) is assumed invertible, with inverse \(\mathbf{h}(\cdot,t)\). The observation region \(\Omega_{\mathrm{obs}}\) consists of the image positions that remain within \(\Omega\) throughout the exposure. Under brightness constancy, the blurred observation \(I_{\mathrm{b}}:\Omega_{\mathrm{obs}}\to\mathbb{R}\) is

\begin{equation}\label{eq:1}
\begin{gathered}
I_{\mathrm{b}}(\mathbf{p})=(\mathcal{B}I_{\mathrm{s}})(\mathbf{p})+N(\mathbf{p}),\\
(\mathcal{B}I_{\mathrm{s}})(\mathbf{p})=\frac{1}{T}\int_0^T I_{\mathrm{s}}\bigl(\mathbf{h}(\mathbf{p},t)\bigr)\,dt.
\end{gathered}
\end{equation}

Here \(N:\Omega_{\mathrm{obs}}\to\mathbb{R}\) denotes additive sensor noise in the same units. For a given motion, the blur operator \(\mathcal{B}\) is linear. This formulation assumes no occlusion, illumination variation or sensor saturation.

We next ask when \(\mathcal{B}\) reduces to a convolution after a change of coordinates. Let \(U\subseteq\mathbb{R}^2\) contain all positions reached by the reference region during the exposure, and let \(\mathbf{W}:U\to\mathbb{R}^2\) be an invertible coordinate mapping. The required condition is

\begin{equation}\label{eq:2}
\begin{gathered}\mathbf{W}\bigl(\mathbf{g}(\mathbf{p}_0,t)\bigr)=\mathbf{W}(\mathbf{p}_0)+\mathbf{v}(t),\\ \mathbf{p}_0\in\Omega,\quad t\in[0,T].\end{gathered}
\end{equation}

Here \(\mathbf{v}:[0,T]\to\mathbb{R}^2\) is a measurable displacement function with \(\mathbf{v}(t_{\mathrm{ref}})=\mathbf{0}\). In these transformed coordinates, every point undergoes the same displacement. Write \(\mathbf{q}=\mathbf{W}(\mathbf{p})\), and denote images transported to these coordinates with a tilde, such as \(\widetilde I_{\mathrm{s}}=I_{\mathrm{s}}\circ\mathbf{W}^{-1}\).

The convolution kernel records the fraction of exposure time spent at each displacement. For a Borel set \(E\subseteq\mathbb{R}^2\), let \(\mathbb{1}_E\) equal 1 inside \(E\) and 0 outside. The dwell measure \(\mu\) assigns to \(E\) the fraction of exposure time during which the displacement lies in \(E\):

\begin{equation}\label{eq:3}
\mu(E)=\frac{1}{T}\int_0^T\mathbb{1}_E\bigl(\mathbf{v}(t)\bigr)\,dt,\qquad \mu(\mathbb{R}^2)=1.
\end{equation}

Substituting this condition into the exposure integral yields

\begin{equation}\label{eq:4}
\begin{gathered}
\widetilde I_{\mathrm{b}}(\mathbf{q})=(\mu*\widetilde I_{\mathrm{s}})(\mathbf{q})+\widetilde N(\mathbf{q}),\\
(\mu*\widetilde I_{\mathrm{s}})(\mathbf{q})=\int_{\mathbb{R}^2}\widetilde I_{\mathrm{s}}(\mathbf{q}-\mathbf{z})\,d\mu(\mathbf{z}).
\end{gathered}
\end{equation}

Here \(\mathbf{z}\in\mathbb{R}^2\) is a displacement vector, and \(d\mu\) denotes integration with respect to the measure \(\mu\). The resulting convolution is linear and obeys the Fourier convolution theorem \cite{rudin2011}. The blur is therefore determined by the geometric mapping \(\mathbf{W}\) and the kernel \(\mu\), and depends only on the time spent at each displacement rather than the temporal order of visits. Because the trajectory is a curve, \(\mu\) generally has no planar density.

Such a common coordinate mapping exists only if the position mappings at any two instants commute, and it can still fail where the motion has fixed points. Rotations and scalings about a fixed centre commute and become translations in log-polar coordinates, whereas an exposure with a moving rotation centre admits no common mapping. When no common mapping exists, the general exposure integral still holds, but the blur no longer reduces to a single convolution.

\subsection{Where information is lost, and what can be recovered}\label{where-information-is-lost-and-what-can-be-recovered}

Suppose \(\mathbf{W}\) and \(\mu\) are known. Because the coordinate mapping is invertible, it loses no information; information loss arises solely from kernel attenuation and sensor noise. Let \(\boldsymbol{\omega}\in\mathbb{R}^2\) be a spatial frequency in the coordinates of \(\mathbf{W}\), and let \(\mathcal{F}\) denote the Fourier transform of a zero-extended image. The transfer function \(K_\mu:\mathbb{R}^2\to\mathbb{C}\) of the kernel is

\begin{equation}\label{eq:5}
\begin{gathered}K_\mu(\boldsymbol{\omega})=\int_{\mathbb{R}^2} e^{-2\pi i\,\boldsymbol{\omega}^{\top}\mathbf{z}}\,d\mu(\mathbf{z})\\ {}=\frac{1}{T}\int_0^T e^{-2\pi i\,\boldsymbol{\omega}^{\top}\mathbf{v}(t)}\,dt,\end{gathered}
\end{equation}

with \(K_\mu(\mathbf{0})=1\) and \(|K_\mu(\boldsymbol{\omega})|\le1\). Blurring multiplies each spatial frequency of the sharp image by \(K_\mu(\boldsymbol{\omega})\).

Two further ambiguities leave the blurred image unchanged. First, a displacement trajectory and its time reversal define the same dwell measure \(\mu\) and hence the same photograph, which constrains how trajectory labels can be defined for supervised learning. Second, a card at twice the distance moving at twice the speed produces identical image-plane trajectories \cite{hartley2004}; we therefore parameterise the operator directly in the image plane.

When the operator is known, regularised least squares with weight \(\lambda>0\) recovers \(\hat I_\lambda=(\mathcal{B}^{*}\mathcal{B}+\lambda\,\mathrm{Id})^{-1}\mathcal{B}^{*}I_{\mathrm{b}}\), where \(\mathcal{B}^{*}\) is the adjoint of \(\mathcal{B}\) and \(\mathrm{Id}\) is the identity. In the coordinates of \(\mathbf{W}\), neglecting boundary effects of the observed region, the inversion acts on each frequency independently:

\begin{equation}\label{eq:6}
\begin{gathered}
\mathcal{F}[\hat I_\lambda](\boldsymbol{\omega})=g(\boldsymbol{\omega})\,\mathcal{F}[I_{\mathrm{s}}](\boldsymbol{\omega})+\frac{\overline{K_\mu(\boldsymbol{\omega})}}{|K_\mu(\boldsymbol{\omega})|^2+\lambda}\,\mathcal{F}[N](\boldsymbol{\omega}),\\
g(\boldsymbol{\omega})=\frac{|K_\mu(\boldsymbol{\omega})|^2}{|K_\mu(\boldsymbol{\omega})|^2+\lambda}.
\end{gathered}
\end{equation}

Here the overline denotes complex conjugation, and \(g:\mathbb{R}^2\to[0,1]\) acts as a frequency gate. For additive noise of standard deviation \(\sigma>0\), frequencies with \(|K_\mu(\boldsymbol{\omega})|^2\gg\lambda\) are recovered with an error of order \(\sigma/|K_\mu(\boldsymbol{\omega})|\), whereas frequencies with \(|K_\mu(\boldsymbol{\omega})|^2\ll\lambda\) are suppressed. Notably, this linear inversion relies neither on the symbolic structure of the QR code nor on a learned image prior.

Decoding a QR code does not require exact pixel-wise reconstruction. While severe blur across the finder patterns or format regions prevents localisation or reading, module errors in the data region are corrected up to the Reed--Solomon capacity of each block \cite{iso18004}. Because no closed-form image metric captures this decoding threshold, we use standard decoders directly to measure how far the assumed operator may depart from the truth.

\subsection{What counts as success, and what the other scores are for}\label{what-counts-as-success-and-what-the-other-scores-are-for}

Let \(m\) be the reference payload registered for a photograph, and let \(\mathsf{R}\) be the read-out operator mapping an image to a decoded string or to the empty string. A recovered image \(\hat I\) yields an exact match if \(\mathsf{R}(\hat I)=m\), an accepted wrong string if \(\mathsf{R}(\hat I)\) is non-empty and differs from \(m\), and a rejection if \(\mathsf{R}(\hat I)\) is empty.

During recovery, when the ground-truth payload \(m\) is unknown, each stage relies on a stage-specific proxy objective. The neural networks are trained on synthetic ground-truth corner positions and trajectories. Candidate trajectories in the neighbourhood search are ranked by the known-module error, defined as the fraction of fixed QR function modules misclassified in the recovered image. Trajectory refinement minimizes the forward reprojection residual \(\lVert\mathcal{B}_\theta\hat L-I_{\mathrm{b}}\rVert^2\), where \(\theta\) denotes a candidate trajectory, \(\mathcal{B}_\theta\) its exposure operator and \(\hat L\) the current binary estimate of the sharp code. Finally, decoded strings are accepted by agreement across multiple decoders.

\subsection{Short exposures of a rigid plane}\label{short-exposures-of-a-rigid-plane}

Most QR codes in practice are printed on a rigid planar surface, which is the focus of this study. Camera motion and target motion combine into a single relative rigid trajectory. Consequently, a part moving past a fixed camera and a camera swept past a stationary card obey the same exposure operator. At each instant, the plane is mapped to the image by a homography, represented by a \(3\times3\) matrix acting on homogeneous coordinates \cite{hartley2004}. Over a short exposure, this relative rigid motion is well approximated by a twist---a rotation about a fixed axis combined with a translation along a fixed direction---traversed at a time-varying speed. A dimensionless progress function \(s:[0,T]\to\mathbb{R}\) with \(s(t_{\mathrm{ref}})=0\) describes the traversal along the twist.

Write \(\overline{\mathbf{p}}_0\) for the homogeneous coordinates of \(\mathbf{p}_0\) and \(\pi\) for perspective projection onto the image plane. We model the image-plane motion as the one-parameter group of homographies generated by the twist,

\begin{equation}\label{eq:7}
\mathbf{g}(\mathbf{p}_0,t)=\pi\!\left(\exp\bigl(s(t)\mathbf{A}\bigr)\overline{\mathbf{p}}_0\right),
\end{equation}

where \(\mathbf{A}\in\mathbb{R}^{3\times3}\) is a trace-free generator with eight dimensionless degrees of freedom determined by the twist, the camera intrinsics and the plane pose at \(t_{\mathrm{ref}}\). Because matrices generated by the same \(\mathbf{A}\) commute, the commutation condition of Section III-A is satisfied; wherever the velocity field of \(\mathbf{A}\) has no fixed point on the code, a flow-box coordinate chart serves as \(\mathbf{W}\) and the blur kernel reduces to the one-dimensional dwell measure of \(s(t)\). Because the plane pose evolves during exposure, the true homography trajectory departs from \(\exp(s\mathbf{A})\) only at second order in the pose change. Lateral translations, axial approaches, in-plane rolls and out-of-plane tilts all arise as special cases.

This formulation offers three practical advantages. First, the motion unknowns reduce to the eight parameters of \(\mathbf{A}\), a one-dimensional speed profile and the reference pose of the code.

Second, the resulting exposure operator is explicit and linear. Dividing the exposure into \(J\) equal time intervals with sample progress values \(s_i\) and homographies \(\mathbf{H}_i=\exp(s_i\mathbf{A})\in\mathbb{R}^{3\times3}\), temporal quadrature gives

\begin{equation}\label{eq:8}
(\mathcal{B}I_{\mathrm{s}})(\mathbf{p})=\frac{1}{J}\sum_{i=1}^{J}I_{\mathrm{s}}\!\left(\pi\bigl(\mathbf{H}_i^{-1}\overline{\mathbf{p}}\bigr)\right).
\end{equation}

This sum evaluates the spatially varying blur directly without explicitly constructing \(\mathbf{W}\) or a two-dimensional kernel array, and uniform weights are exact whenever progress is proportional to exposure time.

\begin{figure}
\centering
\includegraphics[width=\linewidth]{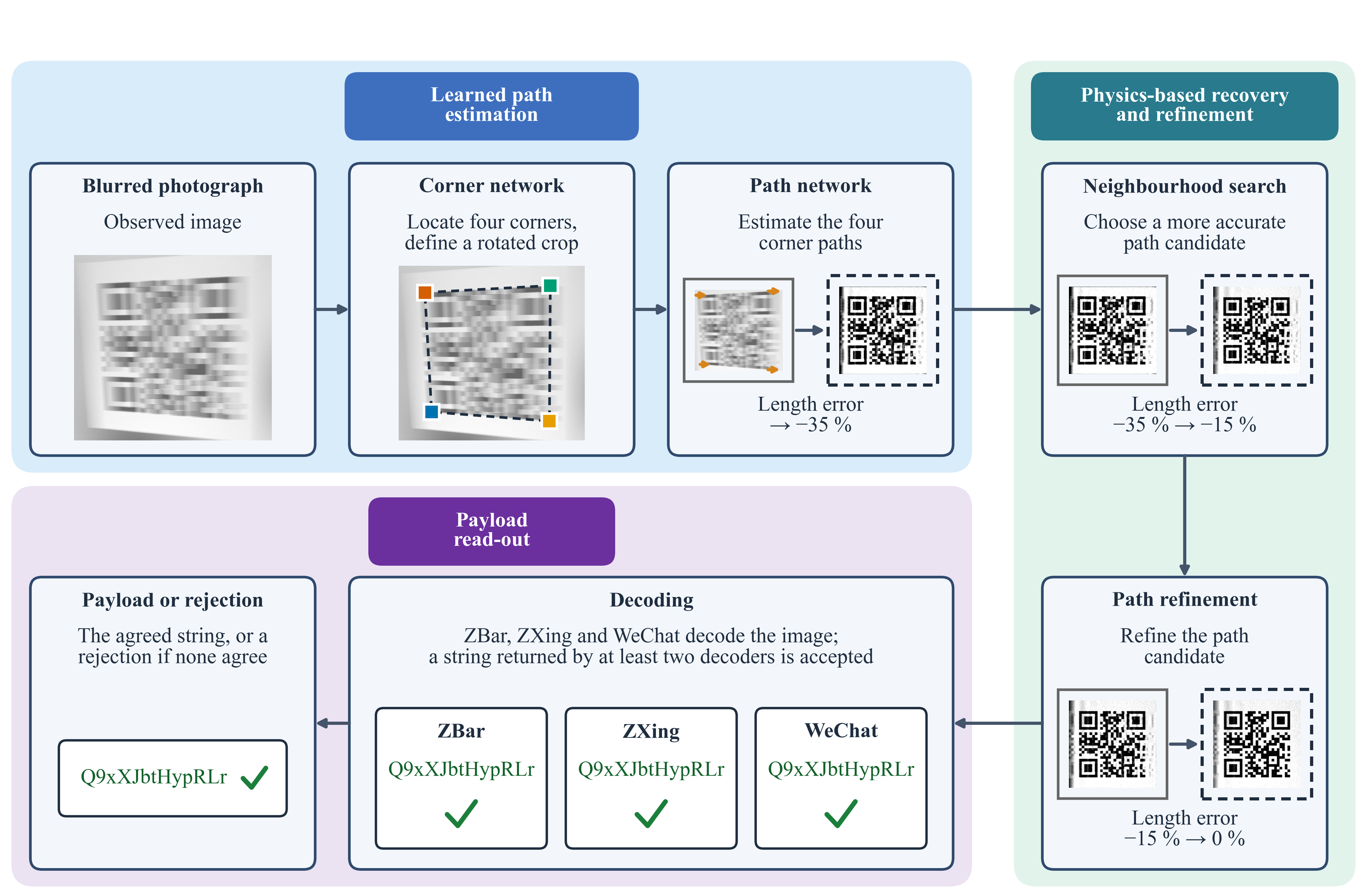}
\caption{Main steps for blurred QR code restoration. A key advantage of this architecture is the tolerance-guided optimisation of candidate motion paths in Stage 2. To our knowledge, this capability is absent from most existing model-based methods and cannot be directly performed by end-to-end black-box networks, yet it substantially improves restoration accuracy.}
\end{figure}

Third, the homography trajectory is compactly parameterised by the paths of the four code corners. Let \(\mathbf{P}_k(s)\in\mathbb{R}^2\) be the image position of corner \(k\in\{1,2,3,4\}\) at progress \(s\in[-1,1]\), in pixels. Expanding to second order in \(s\) gives

\begin{equation}\label{eq:9}
\mathbf{P}_k(s)=\mathbf{P}_k+s\,\mathbf{D}_k+\tfrac12 s^2\mathbf{Q}_k.
\end{equation}

Here \(\mathbf{P}_k\in\mathbb{R}^2\) is the corner position at mid-exposure, \(\mathbf{D}_k\in\mathbb{R}^2\) is its half-displacement and \(\mathbf{Q}_k\in\mathbb{R}^2\) captures the second-order term. Each sample homography \(\mathbf{H}_i\) is uniquely determined by the four corner positions at \(s_i\). Perturbations in blur length, direction, motion centre or speed profile all act through these four corner trajectories. The quadratic term \(\mathbf{Q}_k\) captures both path curvature and monotonic acceleration or deceleration, whereas non-monotonic speed profiles that are slow at both ends and fast in the middle are not considered here.

\section{Method}\label{method}

\subsection{Overall pipeline}\label{overall-pipeline}

As illustrated in Fig. 2, the method takes one motion-blurred photograph and returns either a decoded payload or an explicit rejection. First, a localisation network predicts the four code corners, and a path network predicts the corner trajectories. Each candidate path defines an exposure operator, and regularised linear inversion recovers a candidate sharp image. Next, a neighbourhood search selects a more accurate candidate, followed by three rounds of alternating optimisation to further refine the path. Each recovered image is decoded as soon as it is produced. The run terminates at the first stage where at least two decoders return the same string; if none does, the photograph is rejected.

\subsection{Geometry and motion estimation}\label{geometry-and-motion-estimation}

The greyscale photograph is downsampled by a factor of two, and a heatmap network predicts the four code corners. The photograph is then rotated about the corner centroid so that the code sides lie near the image axes, and an expanded square around the rotated corners defines the input window for the path network. The path network is applied twice: its first pass refines the corners and window, and its second pass predicts the final path, mapped back to photograph coordinates.

For each corner, the network predicts an encoding of \(\mathbf{P}_k\), \(\mathbf{D}_k\) and \(\mathbf{Q}_k\) within the rotated window. Corners are ordered clockwise on the mid-exposure quadrilateral, independent of the finder patterns and of time direction. Training labels describe the image-plane projection of the rendered three-dimensional motion, because depth and three-dimensional velocity are not identifiable from a single photograph.

Time-reversal symmetry imposes an additional constraint: reversing time negates every \(\mathbf{D}_k\) while leaving \(\mathbf{P}_k\), \(\mathbf{Q}_k\) and the photograph unchanged. Let \(\boldsymbol{\delta}\in\mathbb{R}^8\) stack \(\mathbf{D}_1,\dots,\mathbf{D}_4\). The target label stores the symmetric matrix \(\mathbf{M}=\boldsymbol{\delta}\boldsymbol{\delta}^{\top}/\lVert\boldsymbol{\delta}\rVert\in\mathbb{R}^{8\times8}\), which is invariant when \(\boldsymbol{\delta}\) changes sign. Its single nonzero eigenvalue is \(\lVert\boldsymbol{\delta}\rVert\) with eigenvector \(\boldsymbol{\delta}/\lVert\boldsymbol{\delta}\rVert\), so decoding recovers \(\boldsymbol{\delta}\) up to sign. The 36 entries of \(\mathbf{M}\) on and above the diagonal, eight position entries and eight second-order entries form a label in \(\mathbb{R}^{52}\).

Both networks are trained only on synthetic renders of a rigid planar code moving in three dimensions. The renders span all four error-correction levels, quiet zones of zero to four modules, and lateral, approaching, rolling and bounded tilting motion. Illumination variation, sensor noise, gamma encoding and JPEG compression are applied after linear-intensity exposure. Network architectures, training objectives and synthetic data generation are detailed in Sections S5 and S6 of the supplementary material.

\subsection{Physics-based recovery, refinement and read-out}\label{physics-based-recovery-refinement-and-read-out}

The initial network prediction can still fall outside the decoding tolerance when blur is severe, so a neighbourhood search is used to bring the trajectory into the readable basin. This search adjusts all four corner paths together by a uniform length scale and a uniform rotation angle. In principle, the spread of the network's initial error determines the required search range, whereas the parameter tolerance determines the permissible step size. Guided by the in-plane error distribution in Section V-C and the tightest tolerances in Section V-B, we first scan the length scale from 0.70 to 1.35 in steps of 0.05, then scan rotation angles within \(\pm 10^\circ\) in steps of \(5^\circ\) at the selected scale, and finally test local length offsets of \(\pm 0.05\) and \(\pm 0.10\) at the selected angle. These steps keep the nearest grid candidate within 2.5\% in length and \(2.5^\circ\) in direction of the true value, while any non-uniform corner deviations are left to the subsequent per-corner refinement. For each candidate path, we recover a sharp image on linearised intensities by conjugate gradients with \(\lambda=10^{-3}\).

To evaluate candidates without the ground-truth payload, we rely on the fixed function patterns of the QR standard, which specify a known subset of black and white modules. The best path candidate is selected as the one that restores these known patterns with the lowest module classification error. Section S3 of the supplementary material gives the exact search schedule and known-module scoring rule.

Starting from the selected candidate, path refinement further adjusts the path parameters over three rounds with the code size and orientation fixed. Each round extracts a binary code from the current recovery, re-blurs it along the path, and updates the path, intensity and illumination parameters so that the re-blurred image best matches the observed photograph.

To accelerate the procedure, early payload read-out is performed on the input photograph, the initial network recovery, the search-selected recovery, and the recovery after each refinement round. If fewer than two decoders agree on a recovery, we retry the inversion with fallback weights \(\lambda \in \{3\times 10^{-2}, 10^{-2}, 3\times 10^{-3}\}\) and stop at the first weight that achieves two-decoder consensus. The entire run terminates as soon as one image yields a string supported by at least two decoders; if none does after all stages, the image is rejected.

\section{Experiments}\label{experiments}

Controlled simulations first quantify how much trajectory error the linear recovery tolerates, and real motion-blurred photographs then evaluate the complete pipeline across four representative relative-motion families.

\subsection{Evaluation protocol and evidence scope}\label{evaluation-protocol-and-evidence-scope}

We classify each outcome as an exact match if the output equals the registered payload, a wrong string if a nonempty accepted output differs from the payload, or a rejection if no string is reported. All compared methods receive only the test photograph.

Every output image is read by three decoders: ZBar, ZXing, and the OpenCV WeChat engine with its classical detector. Standard preprocessing, including Gaussian smoothing and Otsu binarisation, is applied, and rectified outputs have their finder-pattern levels stretched to 0 and 1. All compared methods follow the same two-decoder acceptance rule.

\subsection{Controlled tolerance measurements}\label{controlled-tolerance-measurements}

This experiment measures how far an assumed motion may depart from the ground truth while linear recovery still returns the correct payload. The design crosses four motion families, four amplitudes, three speed profiles and two card orientations, with ten base scenes per combination. The four families are lateral translation, approach or retreat, rotation about the card normal, and rotation about an in-plane axis. Together they cover in-plane travel, axial distance change, roll, and out-of-plane perspective tilt. The exposure-end/start speed ratios are 1, 2.1 and 5.7, and cards are upright or rotated by 45°. One scene per family, amplitude and speed profile receives three additional noise draws, yielding 1,104 images in total. Renders have at least eight pixels per module, four- or six-module quiet zones, initial tilt up to 20°, JPEG quality of 60--95 and noise scale of 0.005--0.04. Illumination varies across scenes and backgrounds are cluttered, with defocus and sharpening disabled. Tolerance is measured on the 1,065 images whose payload is recovered under the true motion.

Each scan perturbs one assumed parameter while holding all others at their true values. Candidate reconstructions are inverted using conjugate gradients with \(\lambda=10^{-3}\) and at most 40 iterations. Starting from the true motion, the scan first tests an outer endpoint. If the endpoint fails, six bisection steps bracket the transition and we report the midpoint; otherwise, the tolerance is recorded at the scan limit. The scan limits are ±80\% for length, ±60° for translation direction, speed ratios from 1/39 to 39, 20 modules for centre offsets, 90° for axis offsets and 0.30 for added shot-noise scale.

\begin{figure*}
\centering
\centering\includegraphics[width=0.90\linewidth]{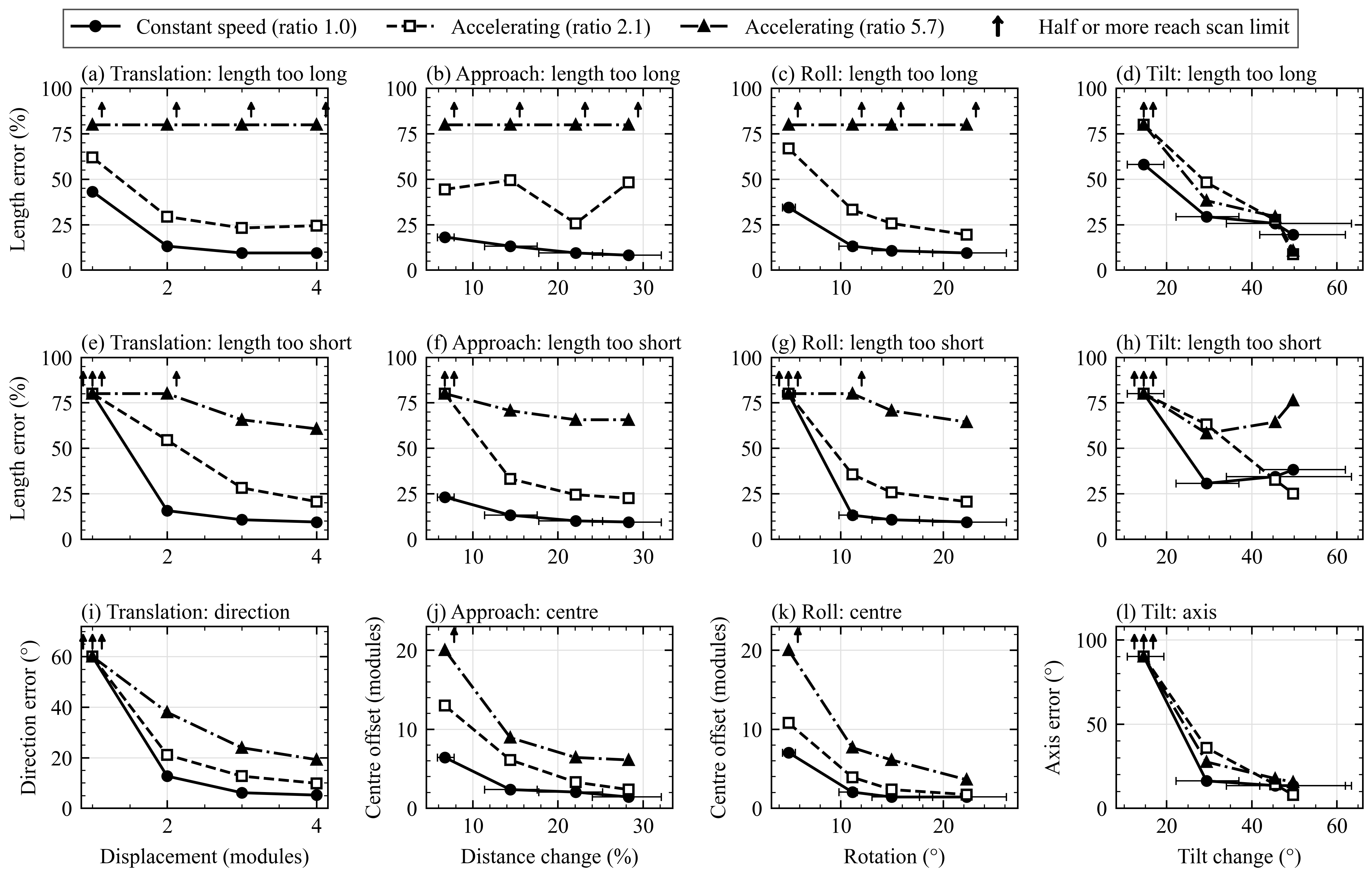}
\caption{Single-parameter motion tolerances across motion families, amplitudes and speed profiles. Both length and geometric tolerances drop sharply from the scan limits as motion grows and are narrowest under constant speed. Across the four families, length underestimation in the middle row remains wider than length overestimation in the top row, especially under large tilt. Panels a--d show length overestimation, panels e--h show length underestimation, and panels i--l show direction, centre or axis error. Solid lines with filled circles, dashed lines with open squares, and dash-dotted lines with triangles plot group medians for speed ratios of 1.0, 2.1 and 5.7, respectively. Horizontal bars indicate the interquartile range of the physical motion amplitude, and upward arrows mark groups where at least half of the images reach the scan limit.}
\end{figure*}

Fig. 3 summarises the single-parameter tolerances across all four motion families and all three speed profiles. First, Fig. 3a--c and 3e--g show that length tolerances for lateral translation, approach and roll tighten as motion grows and are narrowest under constant speed. As the motion reaches 4 modules of translation, 28\% distance change or 22° rotation, the constant-speed tolerance drops from the 80\% scan limit to 8--11\%. At these larger amplitudes, the tolerance rises to 19--54\% at a speed ratio of 2.1, and reaches at least 80\% for overestimation and 61--80\% for underestimation at a ratio of 5.7. Second, comparing Fig. 3e--h with Fig. 3a--d shows that length underestimation is generally tolerated better than overestimation. For example, 2-module translation at a speed ratio of 2.1 tolerates 54\% underestimation against 29\% overestimation, and 50° tilt tolerates 25--76\% underestimation against 9--19\% overestimation. Third, the geometric tolerances in Fig. 3i--l likewise narrow as motion grows and widen with acceleration. Under constant speed, the permitted translation direction error drops from 13° at 2 modules to 5° at 4 modules. Over the same range, the permitted centre offset for approach and roll drops from 2.0--2.3 modules to 1.4 modules, acting mainly by shifting the four corner paths together. For tilt, the permitted axis error similarly tightens from 16°--36° at 29° tilt to 8°--15° at 50° tilt.

Fig. 4a shows that the tolerated range of the assumed speed ratio shifts asymmetrically relative to the true value. Under constant speed, corresponding to a ratio of 1.0, the median lower and upper limits are symmetric on a logarithmic scale at 0.25 and 3.8. Under accelerated motions with speed ratios of 2.1 and 5.7, the upper limit reaches the scan cap of 39 on most images, while the lower limit stays near 0.91 and 1.27. Fig. 4b further shows that simultaneous parameter errors couple tightly for constant-speed lateral translation of 2, 3 and 4 modules, denoted T2, T3 and T4. At T2, a 10\% length overestimate already narrows the direction tolerance from 13° to 8°. Across the larger amplitudes of translation, approach and roll, applying all single-parameter boundary offsets together leaves only 0--13\% of combinations readable under constant speed.

\begin{figure*}
\centering
\includegraphics[width=\linewidth]{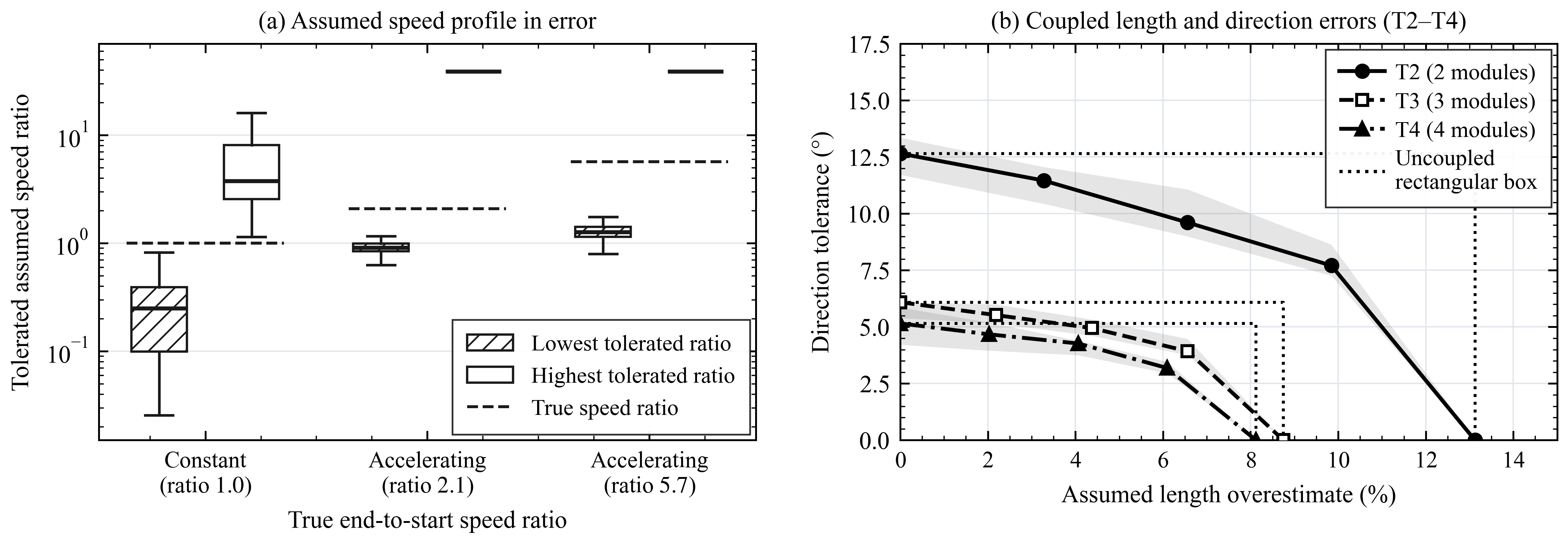}
\caption{Speed-profile tolerance and two-dimensional parameter coupling. In panel a, accelerated motion tolerates a much higher assumed speed ratio than truth, whereas underestimating acceleration quickly causes decoding to fail. In panel b, simultaneous length and direction errors trade off against each other and bend the readable boundary inside the uncoupled rectangular box. Panel a uses hatched and open box plots for the lowest and highest tolerated speed ratios around each dashed true ratio across the three larger motion amplitudes. Panel b plots median curves and shaded interquartile bands of direction tolerance against length overestimation for constant-speed translations T2 through T4.}
\end{figure*}

Finally, we check how sensitive these tolerances are to sensor noise, scene content and card orientation. When the true motion is given, decoding still succeeds under added shot noise with a standard deviation of 13--30\% of full brightness. Repeating the noise draw on the same scene changes the length tolerance by only about 2\%, whereas varying the scene within a motion group changes it by 19\%. Upright cards and cards rotated by \(45^\circ\) yield similar median tolerances. Physical amplitude levels, corner-displacement comparisons, and breakdowns by motion family and card orientation are provided in Section S2 of the supplementary material.

\subsection{Synthetic evaluation of learned estimation}\label{synthetic-evaluation-of-learned-estimation}

We first quantify the initial error of the trajectory network on a synthetic grid that spans the amplitude, speed-profile and geometric levels of Section V-B. Across 4,560 in-plane translation and roll images with motion of at least one module, the initial length ratio spans 0.84 to 1.12 between the 1st and 99th percentiles, with extremes of 0.60 and 1.35. The initial direction error has a median of \(1.2^\circ\), a 99th percentile of \(4.9^\circ\) and a maximum of \(22.6^\circ\). Truncating the search to 0.70--1.35 and \(\pm 10^\circ\) requires 22 inversions and covers 99.9\% of these cases; extending the grid to 0.60--1.35 and \(\pm 20^\circ\) would add 6 inversions for the remaining 0.1\%, whereas narrowing it to 0.80--1.20 and \(\pm 5^\circ\) would save 7 inversions at the cost of lowering in-plane coverage to 98.8\%. Under approach motion with off-centre expansion, the displacement-weighted length ratio has a median of 0.82 and the direction error has a median of \(8.2^\circ\), as the four corners diverge along different rays and rely on the subsequent per-corner fitting stage. Stratified error percentiles across the full design grid are reported in Section S3 of the supplementary material.

We next evaluate how accurately the complete pipeline estimates the motion and recovers the payload on 100 synthetic images not used for training, rendered with perspective distortion, defocus, sensor noise and JPEG compression. Table I compares the cumulative stages of our pipeline against an oracle baseline that uses the ground-truth recorded motion.

\begin{table}[!t]\centering\caption{Decoding error rates and trajectory errors on 100 synthetic images}\label{tab:1}\footnotesize\begin{tabular}{@{}lrr@{}}
\toprule\noalign{}
Stage & Error rate & Shape RMS (modules) \\
\midrule\noalign{}

Direct decoding & 69\% & 0.656 \\
Network trajectory & 27\% & 0.090 \\
Complete pipeline & 23\% & 0.184 \\
Ground-truth trajectory & 17\% & 0.011 \\
\bottomrule\end{tabular}\par\vspace{3pt}\begin{minipage}{\linewidth}\footnotesize A notable observation is that improving a coarse motion estimate brings a large decoding gain, whereas once the estimate is already accurate, further trajectory accuracy no longer translates into better decoding. The first three rows progressively add each stage of our method while retaining earlier decoded results, and the last row replaces the estimated path with the ground-truth quadratic path. Shape RMS is the median root-mean-square error of the assumed trajectory relative to each corner's mid-exposure position, measured in modules; for direct decoding, the assumed trajectory has zero displacement.\end{minipage}\end{table}

Given the ground-truth quadratic path, 83 of the 100 images are recovered, showing that the remaining gap on these images lies in motion estimation rather than in inversion or read-out. Adding search and fitting lowers the error rate from 27\% to 23\%, even though the median trajectory-shape error rises from 0.090 to 0.184 modules. Instead, fitting lowers the median mid-exposure corner error from 0.231 to 0.143 modules. We conjecture that once the trajectory lies within the tolerance basin, refinement helps by correcting the reference position that decoding depends on most, even as the full exposure path becomes slightly less accurate. The 17 images unrecovered even under the recorded motion likely reflect limitations of the motion model, including the quadratic path approximation and unmodelled defocus.

\subsection{Real-photograph evaluation and comparison}\label{real-photograph-evaluation-and-comparison}

\begin{table*}[!t]\centering\caption{Decoding performance on the 955 real motion-blurred photographs}\label{tab:2}\footnotesize\begin{tabular}{@{}
  >{\raggedright\arraybackslash}p{(\linewidth - 10\tabcolsep) * \real{0.1702}}
  >{\raggedright\arraybackslash}p{(\linewidth - 10\tabcolsep) * \real{0.1489}}
  >{\raggedright\arraybackslash}p{(\linewidth - 10\tabcolsep) * \real{0.1489}}
  >{\raggedright\arraybackslash}p{(\linewidth - 10\tabcolsep) * \real{0.1489}}
  >{\raggedright\arraybackslash}p{(\linewidth - 10\tabcolsep) * \real{0.1489}}
  >{\raggedright\arraybackslash}p{(\linewidth - 10\tabcolsep) * \real{0.2340}}@{}}
\toprule\noalign{}
\begin{minipage}[b]{\linewidth}\raggedright
Method
\end{minipage} & \begin{minipage}[b]{\linewidth}\raggedright
Lateral (241)
\end{minipage} & \begin{minipage}[b]{\linewidth}\raggedright
Approach (239)
\end{minipage} & \begin{minipage}[b]{\linewidth}\raggedright
Rotation (239)
\end{minipage} & \begin{minipage}[b]{\linewidth}\raggedright
Coupled 3D (236)
\end{minipage} & \begin{minipage}[b]{\linewidth}\raggedright
All (955)
\end{minipage} \\
\midrule\noalign{}

\textbf{Ours} & 228 / 0 / 13 & 209 / 0 / 30 & 223 / 0 / 16 & 229 / 0 / 7 & \textbf{889 / 0 / 66 (93.1\%)} \\
NAFNet-32 & 209 / 0 / 32 & 203 / 0 / 36 & 200 / 0 / 39 & 186 / 0 / 50 & \textbf{798 / 0 / 157 (83.6\%)} \\
MIMO-UNet+ & 216 / 0 / 25 & 205 / 0 / 34 & 195 / 0 / 44 & 166 / 0 / 70 & \textbf{782 / 0 / 173 (81.9\%)} \\
QAMD & 215 / 0 / 26 & 186 / 0 / 53 & 192 / 0 / 47 & 179 / 0 / 57 & \textbf{772 / 0 / 183 (80.8\%)} \\
MSGN & 197 / 0 / 44 & 200 / 0 / 39 & 191 / 0 / 48 & 140 / 0 / 96 & \textbf{728 / 0 / 227 (76.2\%)} \\
EG-Restormer & 195 / 0 / 46 & 170 / 0 / 69 & 171 / 0 / 68 & 168 / 0 / 68 & \textbf{704 / 0 / 251 (73.7\%)} \\
Restormer & 179 / 0 / 62 & 171 / 0 / 68 & 180 / 0 / 59 & 151 / 0 / 85 & \textbf{681 / 0 / 274 (71.3\%)} \\
DeepDeblur & 133 / 0 / 108 & 176 / 0 / 63 & 155 / 0 / 84 & 80 / 0 / 156 & \textbf{544 / 0 / 411 (57.0\%)} \\
J-MKPD & 98 / 0 / 143 & 147 / 0 / 92 & 126 / 0 / 113 & 61 / 0 / 175 & \textbf{432 / 0 / 523 (45.2\%)} \\
DeMoE & 101 / 0 / 140 & 125 / 0 / 114 & 116 / 0 / 123 & 77 / 0 / 159 & \textbf{419 / 0 / 536 (43.9\%)} \\
SEPC & 65 / 0 / 176 & 111 / 0 / 128 & 94 / 0 / 145 & 37 / 0 / 199 & \textbf{307 / 0 / 648 (32.1\%)} \\
DeblurGAN-v2 & 55 / 0 / 186 & 90 / 0 / 149 & 64 / 0 / 175 & 36 / 0 / 200 & \textbf{245 / 0 / 710 (25.7\%)} \\
Deep Idempotent Network & 45 / 0 / 196 & 76 / 0 / 163 & 60 / 0 / 179 & 22 / 0 / 214 & \textbf{203 / 0 / 752 (21.3\%)} \\
Pan et al. & 51 / 0 / 190 & 57 / 0 / 182 & 49 / 0 / 190 & 37 / 0 / 199 & \textbf{194 / 0 / 761 (20.3\%)} \\
Van Gennip et al. & 41 / 0 / 200 & 54 / 0 / 185 & 48 / 0 / 191 & 20 / 0 / 216 & \textbf{163 / 0 / 792 (17.1\%)} \\
\bottomrule\end{tabular}\par\vspace{3pt}\begin{minipage}{\linewidth}\footnotesize Our method achieves the highest recovery rate in every motion category and overall, while explicit trajectory inversion and learned pixel restoration exhibit distinct category strengths across the four motion types. Columns list the four motion categories and the full dataset, with photograph counts in parentheses. Each entry reports exact matches / wrong strings / rejections, with the exact-match percentage in bold parentheses.\end{minipage}\end{table*}

\begin{table}[!t]\centering\caption{Stage-by-stage ablation of our method on the 955 real motion-blurred photographs}\label{tab:3}\footnotesize\begin{tabular}{@{}lrr@{}}
\toprule\noalign{}
Stage & Correct & Error rate \\
\midrule\noalign{}

Direct decoding & 150 & 84.3\% \\
Network trajectory & 738 & 22.7\% \\
Trajectory search & 836 & 12.5\% \\
Alternating fitting & 889 & 6.9\% \\
\bottomrule\end{tabular}\par\vspace{3pt}\begin{minipage}{\linewidth}\footnotesize Network estimation and inversion recover the majority of blurred photographs, while neighbourhood search and alternating fitting eliminate nearly 70\% of the remaining errors. Each row progressively adds one stage to the pipeline while retaining payloads already decoded at earlier stages. Correct counts exact payload matches under the two-decoder acceptance rule, with zero accepted wrong strings at every stage.\end{minipage}\end{table}

\begin{figure*}
\centering
\centering\includegraphics[width=0.84\linewidth]{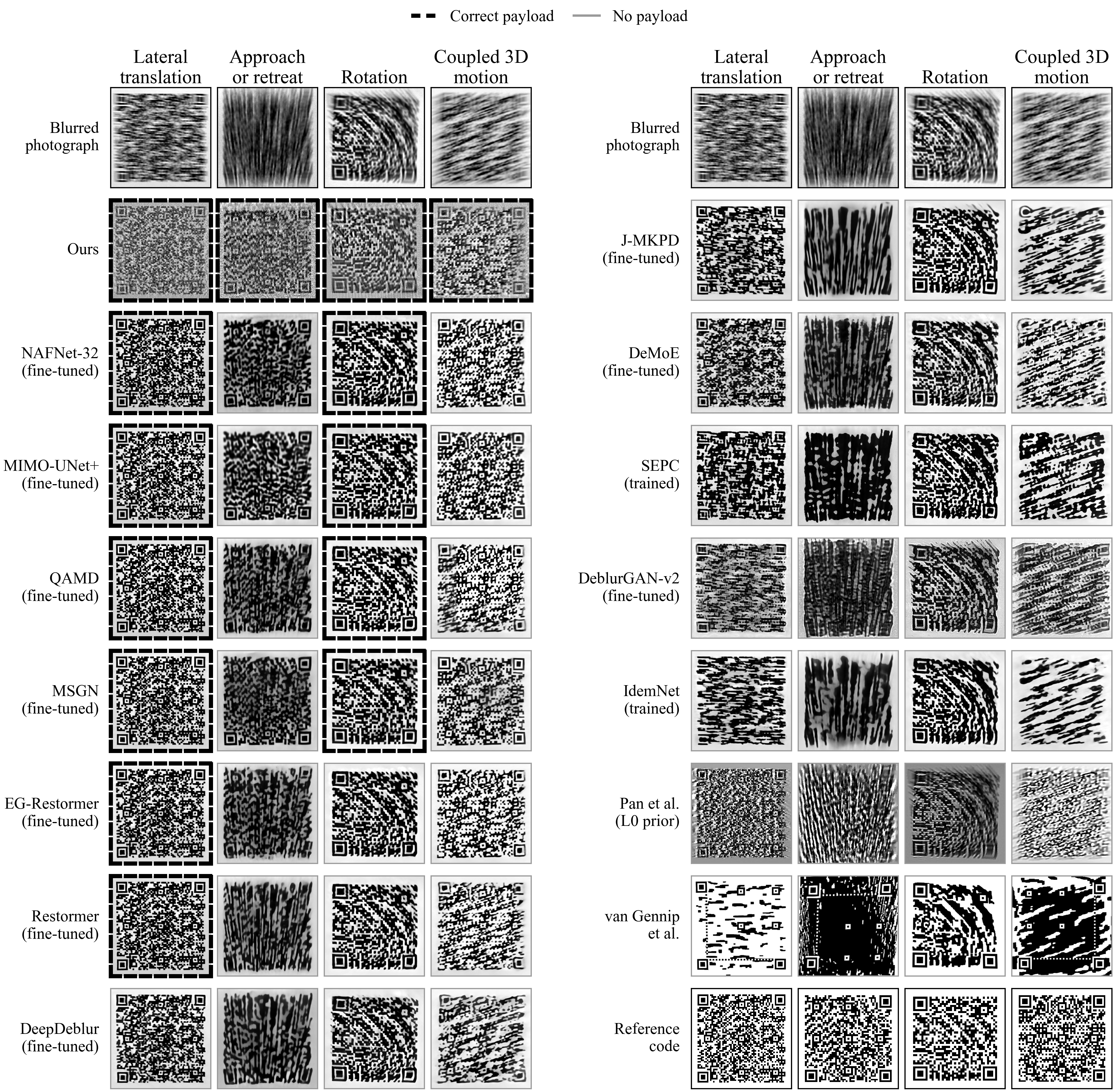}
\caption{Visual comparison on four real motion-blurred photographs across the four motion categories. Several fine-tuned restorers produce visually crisp black-and-white edges yet distort the module grid and fail to decode, whereas our method preserves the decodable module geometry across all four examples. Each panel contains one column per motion category and repeats the blurred input at the top, while the right panel shows the clean reference code at the bottom. The left panel compares our method with the seven strongest baselines from Table II, and the right panel shows the remaining seven baselines in the same order. For visual alignment, all inputs and baseline outputs are rectified with our estimated mid-exposure homography. A thick dashed black border indicates an exact payload match, and a thin solid grey border indicates decoding failure.}
\end{figure*}

Moving from synthetic images to real photographs, we next evaluate our method under genuine optics, sensor noise, and commercial image-processing pipelines. Publicly available datasets of blurred QR codes are scarce, and many suffer from out-of-scope degradations such as optical defocus, missing reference payloads, or tight boundary cropping. By the relative-motion equivalence in Section III-D, sweeping a camera past a stationary card reproduces the same planar exposure operator as a target moving past a fixed camera. Unlike a single mechanical rail confined to one axis and a preset speed profile, operator-driven camera sweeps naturally generate diverse trajectories and acceleration profiles across all four rigid-motion families. Keeping the card stationary also allows a paired static photograph to be taken immediately before each sweep, registering the reference payload by direct decoding without ambiguity. Five operators therefore collected 955 motion-blurred photographs of printed QR cards using four camera models in manual single-frame exposure mode: HUAWEI Mate 80, iQOO 7, OnePlus Ace 6T, and HUAWEI Mate 60 Pro, with two operators sharing a Mate 80. The benchmark covers four relative-motion categories: lateral translation with 241 photographs, axial approach or retreat with 239, rotation about the viewing axis with 239, and coupled 3D motion with 236. Card printing specifications, manual exposure settings, inertial speed recordings, public-dataset comparisons, and the decoding ladder are documented in Section S7 of the supplementary material. The photographs, reference payloads, and inertial recordings are released under CC BY 4.0 \cite{zeng2026data}, and the inference code is released under the MIT License \cite{zeng2026code}.

We compare our method against 12 learned restoration networks---NAFNet-32, MIMO-UNet+, QAMD, MSGN, EG-Restormer, Restormer, DeepDeblur, J-MKPD, DeMoE, SEPC, DeblurGAN-v2 and Deep Idempotent Network \cite{chen2022,cho2021,song2024pr,liu2022tmm,adnet2025,restormer2022,nah2017,carbajal2023,feijoo2025,jiang2024tim,kupyn2019,mao2023tcsvt}---and two classical methods by Pan et al.~and van Gennip et al.~\cite{pan2014,vangennip2015}. Because the learned baselines were not trained on QR images, their released weights recover only 12.1\% to 25.1\% of the photographs, such as 12.1\% for NAFNet-32, 25.1\% for Restormer and 21.4\% for DeMoE. For a fair comparison, we train all 12 networks on 3,000 sharp and blurred image pairs from our synthetic generator, using training payloads disjoint from the photographed cards. Ten networks are fine-tuned from their released deblurring weights, whereas SEPC and Deep Idempotent Network are trained from initialisation with their official losses because SEPC does not release deblurring weights and the repository of Deep Idempotent Network does not provide a checkpoint for its stage-four architecture. The two classical methods require no training. Model checkpoints and classical hyperparameters are selected by decoding rate on 100 synthetic validation images, without access to any test photograph. Every method receives the same 2× downsampled greyscale photograph and is evaluated under the same read-out and acceptance rule. Baseline configurations, runtimes and released-weight results are reported in Section S4 of the supplementary material.

Overall, our method lowers the decoding error rate from 16.4\% for the strongest baseline, NAFNet-32, to 6.9\%, a relative error reduction of 58.0\%. In terms of unread frames that require manual fallback, direct decoding leaves 805 of the 955 photographs unread. NAFNet-32 reduces this unread remainder to 157, and our pipeline reduces it further to 66 while accepting zero wrong strings. Our method recovers 104 photographs rejected by NAFNet-32, whereas NAFNet-32 recovers 13 rejected by ours; the corresponding counts are 123 versus 16 for MIMO-UNet+, 128 versus 11 for QAMD, 170 versus nine for MSGN, 192 versus seven for EG-Restormer, 213 versus five for Restormer and 470 versus zero for DeMoE.

Across the four motion categories in Table II, different method families show complementary strengths. Our method achieves its highest recovery rate on coupled 3D motion at 97.0\%, where complex multi-axis trajectories cause the sharpest drop for all learned pixel restorers. Conversely, our lowest recovery rate occurs on approach or retreat at 87.4\%, which accounts for 30 of our 66 rejections, whereas several learned and classical baselines achieve their best or near-best results in this category. Together with the 13 photographs recovered only by NAFNet-32, this contrast suggests potential for combining explicit trajectory inversion with learned pixel restoration. In addition, photographs below 12 pixels per module at the processed resolution account for 562 of the 955 images but 63 of our 66 rejections, indicating an empirical association with low spatial resolution. Across the four camera models, our recovery rates range from 88.9\% to 94.5\%.

Table III breaks down the contribution of each pipeline stage on the 955 photographs. Direct decoding recovers only 150 photographs, corresponding to an error rate of 84.3\%. Adding corner localisation, network trajectory estimation and inversion recovers 588 additional photographs and drops the error rate to 22.7\%. For photographs still unrecovered, neighbourhood search over motion length and direction rescues another 98 photographs, lowering the error rate to 12.5\%. Finally, three rounds of alternating image-and-path fitting recover 53 more hard cases and bring the final error rate down to 6.9\%. Because the pipeline stops as soon as two decoders agree, search and fitting run only on photographs rejected by earlier stages, accounting for 9\% and 38\% of total processing time, respectively. On a single shared GPU, our complete pipeline takes a median of 5.8 s per photograph including decoding, with a 90th percentile of 33.8 s, compared with 1.2 to 5.9 s for the 12 learned networks and van Gennip et al., and 34 s for Pan et al.

Fig. 5 illustrates representative recoveries across the four motion categories, showing one photograph per category that our method decodes and at least one fine-tuned restorer rejects. Notably, several baseline outputs appear visually sharp and high-contrast yet still fail to decode, whereas our physically inverted reconstructions preserve the module structure required for payload recovery. Moreover, because every intermediate trajectory, reconstructed image and decoder output is explicitly recorded, any failure can be traced to the exact stage where recovery breaks down.

\section{Limitations and Future Work}\label{limitations-and-future-work}

Our current formulation models a planar rigid target under quadratic corner trajectories with uniform temporal weights. While our real-photograph benchmark captures the four primary rigid-motion families, field deployments also encounter curved surfaces, worn labels, partial occlusion, and coupled optical defocus. Extending the physical exposure model to these degradations and combining trajectory inversion with learned pixel restoration are natural next steps.

Furthermore, our synthetic diagnostics show that whole-trajectory accuracy and payload recovery do not always move together. This observation motivates designing refinement objectives guided directly by module decodability rather than geometric path error alone. Finally, our pipeline takes a median of 5.8 s per photograph on a single GPU, spent mostly in repeated inversion and decoding on frames where direct decoding fails. Accelerating the linear solver and candidate screening will enable low-latency deployment on embedded and mobile devices.

\section{Conclusion}\label{conclusion}

This paper investigated how accurately relative rigid motion must be estimated to recover the payload of a motion-blurred QR code under continuous illumination. Within a unified exposure framework, controlled perturbation experiments showed that motion-parameter tolerances are not fixed constants. Tolerances shrink rapidly as blur amplitude grows, and they are tightest under uniform speed yet substantially wider under acceleration. Moreover, length tolerance is strongly asymmetric between underestimation and overestimation, and simultaneous parameter errors interact to shrink the readable basin further.

Guided by these laws, we built an inspectable grey-box recovery pipeline that pairs compact trajectory networks with tolerance-guided candidate search, alternating fitting, and regularised linear inversion. On 955 real motion-blurred photographs spanning four camera models and four rigid-motion families, our pipeline lowers the decoding error rate from 16.4\% for the strongest fine-tuned network to 6.9\%. This corresponds to a 58.0\% relative error reduction, cutting the unread frames from 157 to 66 with zero wrong strings accepted. These results underscore the value of evaluating restoration by machine recognition effectiveness, showing that neither whole-trajectory geometric accuracy nor visual sharpness alone guarantees decodability.

\section*{Acknowledgment}\label{acknowledgment}
\addcontentsline{toc}{section}{Acknowledgment}

The authors thank Prof.~Leida Li of Xidian University for insightful discussions on task-driven quality assessment and machine recognition effectiveness.

\bibliographystyle{IEEEtran}
\bibliography{refs}
\end{document}